\RequirePackage{fix-cm}
\documentclass[twocolumn,epjc3]{svjour3}
\smartqed  
\RequirePackage{graphicx}
\journalname{Eur. Phys. J. C}
\begin{document}

\title{Cylindrical solutions in Mimetic gravity
}
\subtitle{}


\author{Davood Momeni\thanksref{e1,addr1}
        \and
        Kairat Myrzakulov\thanksref{addr1}
        \and
         Ratbay Myrzakulov\thanksref{addr1}
         \and
         Muhammad Raza\thanksref{addr2}
}

\thankstext{e1}{e-mail: davoodmomeni78@gmail.com}


\institute{Eurasian International Center for Theoretical Physics and Department of
General \& Theoretical Physics, Eurasian National University, Astana 010008, Kazakhstan \label{addr1}
           \and
           Department of Mathematics, COMSATS Institute of Information Technology, Sahiwal 57000, Pakistan \label{addr2}
}

\date{Received: 12-May-2015 / Accepted: 17-May-2016}

\maketitle

\begin{abstract}
This paper is devoted to investigate  cylindrical solutions  in  mimetic gravity. The explicit forms of the metric of this theory, namely mimetic-Kasner (say) have been obtained. In this study we have noticed that the Kasner's family of exact  solutions  needs to be reconsidered under this type of modified gravity. A no-go theorem is proposed for the exact solutions in the presence of a cosmological constant.
\keywords{mimetic gravity \and cosmological constant \and cylindrical solutions}
\end{abstract}

\section{Introduction}
\label{intro}
Exact solutions play a crucial role in general relativity (GR) and modified gravity. The thermodynamic  and dynamics  of the gravitational model are often attributed to the presence of an exact solution, which solves the equations of motion. An introduction of a new technique to find the solutions is also an important and natural way to build modify gravity models \cite{exact-sol}.\par
In cosmology of early Universe, we investigate the generally accepted doctrine that the universe is affecting to what we termed as "topological defects"  through exhaustion of all sources of matter, and suggest that by virtue of a cosmic string mechanism which maintains its available energy is self-gravitating. Energy is being "degraded" in objects which are in the cosmos, but "elevated" or raised to a higher level in strings \cite{topological_defects1,topological_defects2}. One main motivation for us to study exact cylindrical solutions in gravitational theories is to describe such topological defects by Riemannian geometry. A simple description of the above topological defects is to find the cylindrical solution by solving highly non linear field equations.  In GR, the simplest cylindrical model described by the  class of exact cylindrical solutions were found by Kasner and later on studied by several authors \cite{Kasner1}
-\cite{Tian}.

This type of cylindrical solution has remained popular in the literature for some time with the name of cosmic string as a model to describe topological defects of early cosmology and closed time like curves.

By definition, the time independent cosmic string  metric has the following properties:

\begin{eqnarray}\label{rho'}
&&g_{\mu\nu}(t,r,\varphi,z)=
\left\{
\begin{array}{lr}
g_{ta}=0\ , & a=\{r,\varphi,z\} \\
\partial_{t}g_{\mu\nu}=0
\ , & \mu,\nu=\{t,r,\varphi,z\}
\\
\partial_z,\partial\varphi
\ , & symmetries
\\
\mathcal{R}^3\times \mathcal{S}^1
\ , & Topology%
\end{array}%
\right. \ .
\end{eqnarray}%

A cosmic string with cylindrical symmetry describes an exact solution in GR in which a line source can be represented by an interior solution in the limit where its radius tends to zero.

Modified gravity has received much attention in recent years due to its interesting properties, which offer important solutions to cosmological queries about the origin of the Universe \cite{obs1}-\cite{obs3}. Modified gravity
is an alternative theory for gravity obtained from action principle and is produced by the replacement of the Einstein-Hilbert action with general function of the curvature and higher derivative terms of it (see \cite{Nojiri:2010wj}-\cite{ijgmmp} for reviews).  It has many possible uses in the gravitational physics and has also been investigated as a potential candidate for the formation of cosmic strings. For example cosmic strings investigated
 in $f(R)$ gravity \cite{f(R)_strings}-\cite{f(R)_strings*}, teleparallel theories \cite{Torsion_strings}-\cite{Houndjo:2012sz}, brane worlds \cite{Braneworld_strings}, Kaluza-Klein
models \cite{KK_strings}, Lovelock Lagrangians\cite{Lovelock_strings}, Gauss-Bonnet
\cite{GB_strings}-\cite{Houndjo:2013us}, Born-Infeld \cite{BI_strings1} -\cite{BI_strings2}, bimetric theories\cite%
{Bimetric_strings}, non-relativistic models of gravity \cite{Misc_strings},  in scalar-tensor theories \cite{ScalarTensor_strings1}- \cite{ScalarTensor_strings7},   Brans-Dicke theory \cite{Delice:2006uz}- \cite{Ciftci:2015cua}, dilation gravity \cite{Dilaton_strings1}-\cite{Dilaton_strings2}, non-minimally coupled models of gravity \cite{Harko:2014axa}
and recently  the  Bose-Einstein condensate strings \cite{Harko:2014pba}.
\par

Recent research has allowed a prescribed number of models to propose, by what is called the "Mimetic Gravity" (MG)  \cite{Chamseddine:2013kea},  which are devoted to resolve the dark matter problem using a class of restricted disformal transformations  $g_{\mu\nu}\to\hat{g}_{\mu\nu}=\Omega(\phi)g_{\mu\nu}$ ($\phi$ is an auxiliary field which can be a complex function) of the physical metric $g_{\mu\nu}$. In the multiple remarks people proposed to follow this main idea from different points of view \cite{Momeni:2014qta}-\cite{Nojiri:2014zqa}.
The general structure of the  cosmic strings in MG has not been  investigated to render any comparison of this structure of strings with that of other classical possible. The subject naturally divides itself into two sections, which we here propose to treat separately; cylindrical solutions in empty space, and passing on to the presence of the non zero cosmological constant. \par
Our plan in this work is as the following: In Sec. (\ref{review}) we briefly  review the basis of mimetic gravity, action and equations of motion. In Sec. (\ref{2}) we prove a possible equivalency between constant Ricci scalar solutions of this theory and Einstein-massless scalar field theory. In Sec. (\ref{eqs}) we study static spacetimes in cylindrical form. In Sec. (\ref{kasner}) we study the possibility to have Kasner solutions as the known solutions in GR in this type of modified gravity. In Sec. (\ref{time}) we study solutions with time dependent scalar field. In Sec. (\ref{lambda}) we investigate extensions of static solutions in the presence of a cosmological constant term. We'll conclude in Sec. (\ref{conclusion}).

\section{Brief review of mimetic gravity}\label{review}
In this section we give a brief sketch of the formalism of the MG theory \cite{Chamseddine:2013kea}.  Let us consider a specific class of disformal transformations and call it reparametrization of the metric $g_{\mu\nu}=(\tilde{g}^{\alpha\beta}
\partial_\alpha\phi\partial_\beta\phi)\tilde{g}_{\mu\nu}$. A further generalization of the Einstein-Hilbert Lagrangian along the disformal transformation leads to so-called mimetic gravity models, whose Lagrangian is the function of the Ricci scalar and the
 scalar field  $\phi$ itself:
\begin{equation}\label{eqn:mg2}
S=-\frac{1}{2}\int d^4x\sqrt{-g(\tilde{g}_{\mu\nu,\phi})}\{R[g_{\mu\nu}(\tilde{g}_{\mu\nu},\phi)]+\mathcal{L}_m\}.
\end{equation}
We had supposed that
 $\tilde{g}_{\mu\nu}$ is an auxiliary metric and $\phi$ denotes an auxiliary (non ghost) field.
Eq. (\ref{eqn:mg2}) is invariant under
the transformation $$g_{\mu\nu}=(\tilde{g}^{\alpha\beta}
\partial_\alpha\phi\partial_\beta\phi)\tilde{g}_{\mu\nu},$$ hence the name disformal-symmetry. The simplest example of the above MG action is the standard Einstein-Hilbert with $\tilde{g}^{\alpha\beta}
\partial_\alpha\phi\partial_\beta\phi=1$. In addition, through the main part of the paper we assume that the metric is static and non-dynamical, therefore the only dynamical variable in the theory is the internal scalar field $\phi$. In this case the equations of motion for the scalar field $\phi$ is nonlinear.  The action given in (\ref{eqn:mg2}) is an alternative reformulation for GR, isolating the conformal degree of freedom of scalar field $\phi$ in a covariant way. This was done by introducing a physical metric $g_{\mu\nu}$ defined in terms of an auxiliary metric $\tilde{g}^{\alpha\beta}$ and a scalar field $\phi$ appearing through its first derivatives $\partial\phi$.

Variation of (\ref{eqn:mg2}) with respect to $g_{\mu\nu},\phi$  gives equations of motion
 (EOM)(see e.g. Ref. \cite{Chamseddine:2013kea}),
\begin{equation}\label{eqn:mg3}
(G^{\mu\nu}-T^{\mu\nu})-(G-T)g^{\mu\alpha}g^{\nu\beta}\partial_\alpha\phi\partial_\beta\phi=0,
\end{equation}
\begin{equation}\label{eqn:mg4}
\frac{1}{\sqrt{-g}}\partial_\kappa[\sqrt{-g}(G-T)g^{\kappa\lambda}\partial_\lambda\phi]=\nabla_\kappa[(G-T)\partial^{\kappa}\phi]=0.
\end{equation}

The resulting EOMs given in Eqs. (\ref{eqn:mg3},\ref{eqn:mg4}) are split into a traceless equation obtained through variation with respect to the auxiliary metric $g_{\mu\nu}$ in Eq. (\ref{eqn:mg3} ) and an additional second order generalized Klein-Gordon equation Eq. (\ref{eqn:mg4}) for the trace part. Consequently the conformal degree of freedom became dynamical even in the absence of matter when $\mathcal{L}_m=0$. One could show that this extra degree of freedom for $\phi$ in the flat cosmological background can mimic cold dark matter.\par

It is interesting to note that the fanciful integration of the second field equation (\ref{eqn:mg4})  over the manifold $\mathcal{M}$ leads to the vanishing normal derivative of the scalar field $n_{\mu}\partial^{\mu}\phi|_{\partial{\mathcal{M}}}=0$, since it occurs in the non GR regime $(G-T)\neq0$:
\begin{eqnarray}
&&\int_{\partial{\mathcal{M}}}d^3x\sqrt{h}(G-T)n_{\mu}\partial^{\mu}\phi =0.
\end{eqnarray}
Regarding the trace of the EOM  (\ref{eqn:mg3}), there is a main difference with GR $(G-T)=0$: the scalar field $\phi$ has a definite form, and is  being expressed in the following:
\begin{eqnarray}
&&g^{\alpha\beta}\partial_\alpha\phi\partial_\beta\phi=1\label{norm}.
\end{eqnarray}
It suggests a solution from $\phi\in \mathcal{C}$ for metric with appropriate signature of the metric $sign(g)=-2$ (In cosmological backgrounds we obtain $\phi\in\mathcal{R}$, due to the isotropicity and homogeneity).

\par
\section{ Notes on constant curvature MG}\label{2}
In this section we prove a general theorem about the exact solutions in MG with the case in which Ricci scalar $R=R_{\mu}^{\mu}$ remains constant. Such type of solutions could be used to explain late time behavior of cosmos in de Sitter epoch as well as solutions with cosmological constant which lead to the Schwarzschild-(Anti) de Sitter spacetime with a wide class of different applications from cosmology to string theory. Calculations show that the conformal degree of freedom can be eliminated by adjusting the constant curvature condition, providing conditions to compare with the exact solutions of MG and GR.

We have discussed here about constant curvature cases and definitely consider "Buchdahl's model" which is a solvable model, and whose action is Einstein-Hibert with a massless scalar field:
\begin{eqnarray}\label{Buchdahl}
&&S=-\frac{1}{2}\int{d^4x\sqrt{-g}\Big(R+\mu\partial_{\alpha}\phi\partial^{\alpha}\phi\Big)}.\label{buchdahl}
\end{eqnarray}
 This action has static (time independent), exact solutions with different symmetries \cite{Buchdahl:1959nk}. We propose the following theorem:
\par
\texttt{Theorem}: \emph{It is universally found that the MG with constant curvature geometries describes the same theory of one  proposed by Buchdahl, These two models are equivalent in their actions and dynamical features.}\\
\texttt{Proof}: When the equation of motion (\ref{eqn:mg3}) is rewritten in the case $R=contant$, a full equivalence of MG with the equations of the motion given in (\ref{buchdahl}) may be observed with $\mu=-\frac{G_{\mu}^{\mu}}{2}$ ; this applies especially the exact solutions. If the equations of motion of the system can be solved with a specific symmetry,  the solutions will also be recovered by those given in \cite{Buchdahl:1959nk}:
\begin{eqnarray}\label{Buchdahl}
&&S=-\frac{1}{2}\int{d^4x\sqrt{-g}\Big(R-\frac{G}{2}\partial_{\mu}\phi\partial^{\mu}\phi\Big)},
\end{eqnarray}
whose axially symmetric solutions are given by the following:
\begin{eqnarray}
ds^2=e^{2(\gamma-\beta\psi)}(-dr^2-dz^2)-r^2e^{-2\beta\psi}d\varphi^2+e^{2\beta\psi}dt^2.
\end{eqnarray}
In which the metric functions $\{\psi,\gamma\}$ satisfy the following equations:
\begin{eqnarray}
&&r^{-1}(r\psi_{,r})_{,r}+\psi_{,zz}=0,\\&&
\gamma(r,z)=\int\Big[(r\psi_{,r}^2-\psi_{,z}^2)dr+2r\psi_{,r}\psi_{,z}dz\Big],\\&&
\phi=2\lambda\psi,\ \ \beta=\pm(1-2\lambda^2)^{1/2},
\end{eqnarray}
where $\psi_{,r},\psi_{,z}$ denote derivatives with respect to $r,z$.
If $\beta=1$ or $\lambda=0$ and $\gamma_{,z}=\psi_{,z}=0$ , Buchdahl solution reduced to the vacuum Levi-Civita cylindrical solution \cite{LC}:
\begin{eqnarray}\label{LC}
ds^2=r^{2  m}dt^2-r^{2m(m-1)}(dr^2+dz^2)-r^{2(1-
m)}d\varphi^2.
\end{eqnarray}

For the dynamical point of view (for example in MG scenario for inflation) it makes no difference between the action of GR with a massless scalar field and the one in MG. This equivalency between constant Ricci scalar $R$ solutions of MG and Einstein-massless scalar field theory is an essential feature of any purely kinetic (only function of $\partial_{\alpha}\phi\partial^{\alpha}\phi$ ) form of this type of disformal deformation of GR.

\section{Field equations for a static cylindrical spacetime} \label{eqs}
To find the solution of field equations in GR or MG than in  vacuum (exterior) or interior (non vacuum) regions, we may have to adopt the more appropriate representation of coordinates for metric. An appropriate standard frame for cylindrical polar coordinates is $(t,r,\varphi,z)$. The most general form of a cylindrically symmetric spacetime is given by the following metric:
\begin{equation}
ds^{2}=A(r)dt^2-dr^2-B(r)d\varphi^2-C(r)dz^2
\label{lineelement}
\end{equation}%

Because to keep generality we write down EOMs in the presence of a non zero cosmological constant $\Lambda$  in  following forms:

\begin{eqnarray}\label{A2}
&&\frac{A'}{A}\frac{C'}{C}+\frac{A'}{A}\frac{B'}{B}+\frac{B'}{B}\frac{C'}{C}-4\Big(\frac{A''}{A}+\frac{B''}{B}+\frac{C''}{C}\Big)\\&&\nonumber+2\Big(\big(\frac{A'}{A}\big)^2+\big(\frac{B'}{B}\big)^2+\big(\frac{C'}{C}\big)^2\big)+12\Lambda=0.
\end{eqnarray}

\begin{eqnarray}
&&\frac{A'}{A}\frac{C'}{C}-\big(\frac{A'}{A}\big)^2-\big(\frac{C'}{C}\big)^2+2\Big(\frac{A''}{A}+\frac{C''}{C}\Big)=4\Lambda\label{B2}.
\end{eqnarray}%

\begin{eqnarray}
&&\frac{B'}{B}\frac{C'}{C}-\big(\frac{B'}{B}\big)^2-\big(\frac{C'}{C}\big)^2+2\Big(\frac{B''}{B}+\frac{C''}{C}\Big)=4\Lambda\label{C2}.
\end{eqnarray}

\begin{eqnarray}
&&\,{\frac {{ A'}\,}{A}}\frac{B'}{B}
-
\big(\frac{B'}{B}\big)^2-\big(\frac{A'}{A}\big)^2+2\Big(\,{\frac {{ A''}}{A}}+\,{\frac {{B''}}{B}}\Big)=4\Lambda\label{D2}.
\end{eqnarray}

respectively. The aim here is to find extended solutions for the system of equations given in (\ref{A2}-\ref{D2}) for $\Lambda=0,\Lambda\neq0$. We suppose that the scalar field is cylindrical (could be time dependent) $\phi=\phi(r,t)$.
In empty space, $\Lambda=0$, the gravitational field equations Eqs.(\ref{A2}-\ref{D2}) gives the following exact solution for scalar field  Eq.(\ref{norm}), 




\begin{eqnarray}
&&\phi(r)=i(r-r_0)\in\mathcal{C}.
\end{eqnarray}
We find that, except for the factor $r_0$, this solution for scalar field is transformed to a complex function in $\mathcal{C}$.

The particular problem is to solve a system of non linear differential equations (\ref{A2}-\ref{D2}) and to find the metric functions $A,B,C$.

\section{ Realization of Kasner's solution}\label{kasner}
Exact solutions for the  Einstein equations with cylindrical symmetry  can lead to the following  two parametric metric, named Kasner solution:
\cite{Kasner1,Kasner2,exact-sol},
\begin{eqnarray}  \label{Kasner1}
ds^{2} = (kr)^{2a}dt^{2} - dr^{2} - \beta^{2} (kr)^{2(b-1)}r^{2}d{\varphi}^2-(kr)^{2c}dz^{2},\nonumber
\end{eqnarray}
where $k$ sets the length scale and $\beta$ is a constant (is related to the
deficit angle of the conical space-time)
\cite{Kasner1,Kasner2,exact-sol}.
Thus, to solve the Einstein  equation, we consider, not merely the value of metric functions for which $R_{\mu\nu}=0$, but the value of $\{a,b,c\}$ for every possible value of the parameters:
\begin{equation}
a + b + c=a^2 + b^2 + c^2=1.  \label{Kasner2}
\end{equation}
For Kasner metric, $R=0$ and for quasi Kasner $R\neq0$, since we supposed that in MG, $(G-T)=-R\neq0$, so the Kasner metric is also a trivial solution in MG .
Does the quasi Kasner solution  with $R\neq 0$ solves our MG system of the equations given by Eqs. (\ref{A2}-\ref{D2}) or not?. We investigate this problem in the following cases:
\par

\begin{itemize}
\item  \it{Quasi-Kasner solutions in MG}: $A=(kr)^{2a},B=\beta^{2}r^{2} (kr)^{2(b-1)}, C=(kr)^{2c}$:
\end{itemize}

The condition $(G-T)\neq0$  wouldn't stop us  from finding the quasi Kasner's solutions  if we wanted to, and it might just make us surprise enough to find something similar to GR. Substituting this value of solutions in Eqs. (\ref{A2}-\ref{D2}), we observe that there are certain values of the parameters $\big(a,b,c\big)$ for which the quasi Kasner is a solution in MG:

\begin{eqnarray}
&&ds^{2} = dt^{2} - dr^{2} - \beta^{2} k^{2}d{\varphi}^2-
(kr)^{2}dz^{2},\ \ ( 0, 0, 1),\\&&
ds^{2} = dt^{2} - dr^{2} - \beta^{2}r^{2}d{\varphi}^2-
dz^{2},\ \   ( 0,1,0),\\&&
ds^{2} = (kr)^{2}dt^{2} - dr^{2} - \beta^{2} k^{2}d{\varphi}^2-
dz^{2},\ \   ( 1, 0, 0).
\end{eqnarray}

Substituting any solution of this group into the metric (\ref{Kasner1}), we obtain metrics of the type Kasner with  $R=0$. We conclude that quasi Kasner solutions don't exist in MG.
\begin{itemize}
\item \it{ Non Kasner type of the exact solutions}:
\end{itemize}
If we want to perform an elimination of $A$, $C$ and $B$ respectively in Eqs. (\ref{A2}-\ref{D2}), we can specify this through use of a lex ranking for the algebraic problem.  We obtain:
\begin{eqnarray}
&&{\it B'''}=-{\frac {{{\it B'}}^{2}{\it B''}-2\,{{\it B''}}^{2}B}{{\it
B'}\,B}}\label{B}
\\&&
{\it C''}=-{\frac {2\,{\it B''}\,B{\it B'}C-{{\it B'}}^{3}C+{\it B''}
\,{B}^{2}{\it C'}}{{\it B'}\,{B}^{2}}}\label{C}
\\&&{\it A'}=-{\frac {{\it B'}\,{\it C'}\,AB  -2\,{{\it B'
}}^{2}AC+4\,{\it B''}B\,AC}{- B{\it C'}
 ^{2}+ B' B C}}\label{A}.
\end{eqnarray}
with a constraint:
\begin{eqnarray}
-B'^2BC'C-B'^3C^2+B'B^2C'^2+2B''B^2BC'\nonumber \\+2B''B'BC^2 = 0.
\end{eqnarray}
Then by solving these differential equations, regarding the three elements $\{A,B,C\}$ as unknown functions, the values of the latter may be computed as follows:
\begin{eqnarray}
&&B(r)=\tilde{C_3}(r-r_0)^{C_1},
\label{Bsol}
\\&&C(r)={ C_4}( r-r_0)^{-\frac{1}{2}{C_1}+1+\frac{1}{2}n}\nonumber\\&&+{C_5}\,
( r-r_0) ^{-\frac{1}{2}{C_1}+1-\frac{1}{2}n},
\label{Csol}
\\ &&A(r)=C_3e^{-C_1\int  dr\frac{ {C_4}({C_1}+\frac{n}{3}-2)
   (r-r_0){}^{\frac{n}{2}}+3 {C_5}({C_1}-\frac{n}{3}-2) ( r-r_0 ){}^{-\frac{n}{2}}}{
{C_4}
   ({C1}-\frac{n}{3}-\frac{2}{3})
   ( r-r_0 ){}^{1+\frac{n}{2}}+{C_5}
   ({C_1}+\frac{n}{3}-\frac{2}{3})
   ( r-r_0 ){}^{1-\frac{n}{2}}} }.
\label{Asol}
\end{eqnarray}

Here $\tilde{C_3}=C_3 (-1)^{C_1},n=\sqrt{-3C_1^2+4C_1+4},C_3>0$ where its  value  is to be bounded as $C_1\in[-\frac{2}{3},2]$ to assure all real values of $n$ and $r\geq C_2$. 
We mention here that this is verified  that the metric functions Eqs. (\ref{Bsol}-\ref{Asol}) to be  exact solutions to the vacuum  field equations (\ref{A2}-\ref{D2})  using the MAPLE  GR{\it Tensor}  package.
\par
 When $n=0$, the metric is explicitly identified by:

\begin{eqnarray}
&&ds^2=\rho^2 d\tilde{t}^2- d\rho^2-\mu  d\varphi^2-  d\tilde{z}^2,\nonumber\\&&\mu=C_4+C_5,
 \rho=r-C_2,\tilde{t}= t\sqrt{\tilde{C_3}},\ \ \tilde{z}=z\sqrt{C_3} \label{sol2}.
\end{eqnarray}
This solution corresponds to the vacuum Levi-Civita  metric presented in Eq. (\ref{LC}) for $m=1$, i.e. the cosmic string,
The range of the angle $\varphi$ does not precisely coincide with the flat metric $(0,2\pi]$, but the geometry is fairly close to that one, where a metric coincide and meet in the exterior, with a deficit angle, for exterior the quantity $1-4\eta=\mu$, here $\eta$ defines the gravitational mass per unit length of the spacetime.\par
Replacing the solutions given by (\ref{Bsol}-\ref{Asol})  into the metric (\ref{lineelement}), we find the following exact cylindrical symmetric solution for MG with $R\neq0$:
\begin{eqnarray}
&&ds^{2}=C_3e^{-C_1\int  dr\frac{ {C_4}({C_1}+\frac{n}{3}-2)
   (r-r_0){}^{\frac{n}{2}}+3 {C_5}({C_1}-\frac{n}{3}-2) ( r-r_0 ){}^{-\frac{n}{2}}}{
{C_4}
   ({C1}-\frac{n}{3}-\frac{2}{3})
   ( r-r_0 ){}^{1+\frac{n}{2}}+{C_5}
   ({C_1}+\frac{n}{3}-\frac{2}{3})
   ( r-r_0 ){}^{1-\frac{n}{2}}} }dt^2\\&&\nonumber-d\rho^2-\tilde{C_3}\rho^{C_1}d\varphi^2-\Big({ C_4}\, \rho ^{-\frac{1}{2}\,{
 C_1}+1+\frac{1}{2}\,n}+{
C_5}\, \rho ^{-\frac{1}{2}\,{ C_1}+1-\frac{1}{2}\,n}\Big)dz^2,
\label{metricexact}
\end{eqnarray}
here the Ricci scalar is found as follows:
\begin{eqnarray}
&&R= \Big[\left( -1 \right) ^{1+\,{C_1}} n^2{
C_1}\,{ C_4}\,{\it C_5}\Big]\rho ^{-1+2\,{C_1}}.
\end{eqnarray}
We observe that the Ricci scalar $R$ given by the above expression, is clearly non zero.  The  Kretschmann scalar $\mathcal{K}=R_{\mu\nu\alpha\beta}R^{\mu\nu\alpha\beta}$ can be calculated for metric given in (\ref{metricexact}) as follows:
\begin{eqnarray}
&&\mathcal{K}=\frac{D(\rho,C_1,..,C_5)\Big({C_4} {\rho}^{n/2}+{ C_5}\,{
\rho}^{-n/2} \Big)^{-1}}{{\rho}^{4} \Big({C_4}\, \left( -3\,{ C_1}+n+2
 \right) {\rho}^{n/2} -  {C_5}\left( 3\,{C_1}+n-2 \right)\,{\rho
}^{-n/2} \Big) ^{4}  }.
\end{eqnarray}
Where we observe that $D(\rho,C_1,..,C_5)$ is a non-singular function for $\rho\in\mathcal{R}$. The  Kretschmann scalar has singularities located at $\rho=0,\rho_{+},\rho_{*},\rho_l$ where:
\begin{eqnarray}
&&\rho_{+}= \left( {\frac {\sqrt {- {
}\,{C_4} \, \left( -2+3\,{C_1}-n \right) }}{\sqrt{ {C_5}\left( 3\,{
C_1}+n-2 \right) }}} \right) ^{-\frac{2}{n}}, C_5>0\label{rho1}\\&&
\rho_{*}=\left( {\frac {\sqrt { {
}\,{C_4} \, \left( -2+3\,{C_1}-n \right) }}{\sqrt{- {C_5}\left( 3\,{
 C_1}+n-2 \right) }}} \right) ^{-\frac{2}{n}}
, C_5<0,\label{rho2}\\&&
\rho_l={{\rm e}^{-\frac{1}{n}\ln  \left( -{\frac {{C_4}}{{C_5}}} \right)
}},\ \  \frac {{C_4}}{{C_5}}<0\label{rho3}.
\end{eqnarray}
Here $\frac{C_4}{C_5}$ are arbitrary constants.  We see that, this spacetime has singularities as in the GR case. It is well known that the static Levi-Civita spacetime is singular at $r=0$, except for $m=0$, $m=\pm\frac{1}{2}$ and $m\to \infty$. For these values of $m$, the solution is regular and flat. \par


\par
\section{Solutions for time dependent scalar field $\phi=\phi(r,t)$}\label{time}
If we had a sufficiently complete cylindrical symmetry for the scalar field $\phi(r)$, we might have to investigate cosmic strings more directly by considering the resultant of the solutions and the non static scalar field $\phi(r,t)$ which moves in the same static cylindrical background.
 If we relax the staticity (time independent) in $\phi$, and we allow it to be time dependent that is $\phi(r,t)$, thus by Eq. (\ref{norm})  of the scalar field, the exact solution is expressed by:
\begin{eqnarray}
&&\phi(r,t)=at+b\pm\int dr\sqrt{\frac{a^2}{A(r)}-1}.
\end{eqnarray}
 The preceding investigation is based upon the assumption that in solving the Hamilton-Jacobi equation  (\ref{norm}) the isotropcity of the field does not valid. The $t-r$ component of  Eq. (\ref{eqn:mg3}) gives us the following solutions:
\begin{itemize}
\item a=0: This case corresponds to the static and cylindrical solutions which we investigated in the previous section.
\item $A(r)=\pm a\neq\{0,\frac{\pm 1}{2}\}$: In this case, the modified forms of the EOMs given in Eqs. (\ref{A2}-\ref{D2}) are obtained by the following system:
\begin{eqnarray}
&&(-a\pm\frac{ 1}{2})\Big[\,{\frac {{ B''}}{B}}+{\frac {{ C''}}{C}}-\frac{1}{2}\Big(\,{\frac {{{ B'}}^{2}}{{B}^{2}}}+\,{\frac {{{ C'}}^{2}}{{C}^{2}}}\Big)\Big]=0,\label{tt-a}
\end{eqnarray}

\begin{eqnarray}
&& B' C'=0,
\label{rr-a}
\end{eqnarray}

\begin{eqnarray}
&&\frac{B''}{B}+\frac{C''}{C}-\frac{1}{2}\Big(\big(\frac{B'}{B}\big)^2+\big(\frac{C'}{C}\big)^2\Big)=0\label{phiphi-a}.
\end{eqnarray}
and

\begin{eqnarray}
&&\frac {{B''}}{B}-\frac{1}{2}\big(\frac{B'}{B}\big)^2=0\label{zz-a}.
\end{eqnarray}
When $B'=0,C'\neq0$ (or vice versa, without disturbing the generality of discussion), we obtain:
\begin{eqnarray}
&&C(r)=\frac{1}{2}\big(\frac{c^2}{2}r^2+dc r+\frac{d^2}{2}\big).
\end{eqnarray}

the metric is obtained as follows:
\begin{eqnarray}
&&ds^2=(\mp a)dt^2-dr^2-B^2d\varphi^2\\&&\nonumber-\frac{dz^2}{2}\big(\frac{c^2}{2}r^2+dc r+\frac{d^2}{2}\big)\label{Bclass}.
\end{eqnarray}
Where $B$ is an arbitrary constant. Another choice is obtained when $B'\neq0,C'=0$:

\begin{eqnarray}
&&ds^2=(\pm a)dt^2-dr^2\\&&\nonumber-\frac{1}{2}\big(\frac{c'^2}{2}r^2+d'c' r+\frac{d'^2}{2}\big)d\varphi^2-C^2dz^2\label{Cclass}.
\end{eqnarray}
here $\{C,d,c,d',c'\}$ denote a new set of parameters and $\pm a>0$ (plus sign for $a\in\mathcal{R}^{+}$ and minus for $a\in\mathcal{R}^{-}$).  It is remarkable that the class (\ref{Bclass}) represents a cosmic string when $d=0$ , and the class (\ref{Cclass}) when we set $d'=0$.

\end{itemize}
Most remarkable in this non static case  is the structure of the spacetime. There are never more than two parameters, and the spacetime, usually so conspicuous in $g_{\varphi\varphi}$, are reduced  to cosmic strings. The reason to study cylindrical solutions with time dependent scalar field backs to a big difference between spherically  and cylindrically symmetric metrics. In  GR We know that according to the Birkhoff theorem, there always exist  a timelike Killing vector $\partial_t$  in the spherically symmetric vacuum solution. Consequently we can say that the spherically symmetric vacuum gravitating system is necessarily static. However, the situation drastically changes when we consider the cylindrically symmetric systems because there is no analogue of Birkhoff’s theorem in cylindrical symmetry. During the gravitational collapse of a cylindrically symmetric system, gravitational waves can be emitted and the exterior region of a collapsing cylindrical body is not static \cite{Delice:2004wk}.  Our solution with time dependent
scalar field with a static metric could be a subset of the most general class of Einstein-Rosen (ER) gravitational wave solutions in mimetic
gravity in comparison with the GR solutions \cite{Ozgur1,Ozugr2}.

\par
\section{Case with cosmological constant}\label{lambda}
In the system of Eqs. (\ref{A2}-\ref{D2}), if we set  $\Lambda\neq0$ and in the absence of the scalar field $\phi=0$ , we already know a generalization of Kasner's solutions to Linet-Tian (LT) family \cite{Linet1,Linet2,Tian} which is written in a slightly different coordinate system as follows:
\begin{eqnarray}\label{LT}
&&ds^2=\Big[\tan(\beta(r+\hat{r}))\Big]^{\gamma_1}
\Big[\sin(2\beta(r+\hat{r}))\Big]^{2/3}dt^2\\&&\nonumber-dr^2-\Big[\tan(\beta(r+\hat{r}))\Big]^{\gamma_2}
\Big[\sin(2\beta(r+\hat{r}))\Big]^{2/3}d\varphi^2\\&&\nonumber-
\Big[\tan(\beta(r+\hat{r}))\Big]^{\gamma_3}
\Big[\sin(2\beta(r+\hat{r}))\Big]^{2/3}dz^2,\\&&\nonumber \hat{r}\in\mathcal{R},\Sigma\gamma_i=0,\Sigma_{i\neq j}\gamma_i\gamma_j=-\frac{4}{3},\gamma_i\in[-\frac{4}{3},\frac{4}{3}].
\end{eqnarray}
 In GR, the LT family was generalized to the case with a massless scalar field $\phi\neq0$  by  the following exact solution\cite{ijmpa}:
\begin{eqnarray}
&&ds^{2}=-dr^2+e^{-2\sqrt{\frac{\Lambda}{3}}r}(\xi^2e^{2\sqrt{3\Lambda}r}+1)^{2/3}\\&&\nonumber\times(dt^2-d\varphi^2-dz^2),\\&&\nonumber \phi(r)=
\pm2 \frac{\sqrt{6}}{3}\tan^{-1}(\xi e^{\sqrt{3\Lambda}r})\label{me}
\end{eqnarray}
here $R=4\Lambda$.
This exact solution has the following properties:
\begin{itemize}
  \item If $\xi^2>0$, the scalar field is real $\phi\in\mathcal{R}$ ,
and the  Kretschmann scalar $\mathcal{K}=R_{\mu\nu\alpha\beta}R^{\mu\nu\alpha\beta}$ is free of any  naked singularity .

\item When  $\xi\in\mathcal{C}$, we have a  naked singularity located at
 $r=r_{0}=-\frac{1}{3}a\ln(|\xi|)$ where $ 0<|\xi|<1$

\item When $|\xi|>1$ the solution is free of naked singularities.

\end{itemize}
 In modified gravity, every simple modified gravity theory admitting constant curvature as solutions should admit this solution as well. This is why LT solution also solves $F(R)$ theory for constant Ricci scalar. But in MG, because we suppose that $(G-T)\neq0$, to obtain different solutions from the GR, consequently this condition is broken, because $G-T=-R-T$, and if  we take $T_{\mu\nu}=\Lambda g_{\mu\nu}$, the constant curvature $R=R_0\neq 4\Lambda$ , as a result LT family doesnt exist in MG. A way to realize the constant curvature solutions in MG is to set $T_{\mu\nu}=0$ and $R=R_0$. But the solution won't be LT solution.

 This argument is principally based on the following general theorem, which is a remarkable extension of \emph{No-go theorems}.\par
{\bf No-go Theorem}:
{\it If we consider the MG with cosmological constant, i.e. the system of differential equations given by  Eqs. (\ref{A2}-\ref{D2}),this theorem clearly state that the solution given by  Eq. (\ref{LT}) is not a solution to the MG theory.}\par
{\bf Proof}:
To obtain the corresponding mathematical proof concerning the general form of equations of motion (\ref{A2}-\ref{D2}) , we eliminate $A,C$ throughout by (\ref{A2}-\ref{D2})
 and obtain (\ref{B}).  If we want to perform an elimination of $A$, $C$, and then $B$ respectively, we can seek this through use of a lex ranking for the algebraic problem.
Paying attention merely to the other equations $[B'C-C'B\neq0 , B'\neq0]$,we see that  there is no solution for the metric functions $\{A,B,C\}$ with cosmological constant. Indeed, the only solution with constant curvature is when $\Lambda=0$ and this solution generally does not coincide with the LT solution.

The general theorem is manifest, and yields a development in any attempt to generalize LT family of corresponding MG Lagrangian.

\section{Discussions and final remarks}\label{conclusion}
In this paper we investigated static cylindrical solutions for mimetic gravity, a conformally invariant version of Einstein gravity with a \emph{non ghost} scalar degree of freedom. Accordingly, if Ricci scalar $R$ be a constant, mimetic gravity reduced to Einstein-Hilbert action with a massless scalar field, (\ref{Buchdahl}). The special limits of the functions of these metrics, namely cosmic strings (say), may be used to investigate generally the forming of a locally flat but globally different cylindrical spacetime exterior to a cosmic string; the actual mass per length could  be determined by computing the metrics in a particular case. Solutions of equations of motion in the vacuum case give  precipitate with three singularities,  when two singular points are coincided.  The summary of results might have been written in a list as:
\begin{itemize}
\item Constant curvature vacuum solutions in mimetic gravity are equivalent to the solutions in Einstein gravity with a massless scalar field.
\item Quasi Kasner solution doesn't exist in mimetic gravity.
\item A family of exact solutions with variable $R$ is found which are different from the Levi-Civita or Kasner family in GR.
This solution has four  Kretschmann's singularities, one is naked singularity on string's axis $\rho=0$, and three cylindrical "horizons" as $\rho_{+}\leq\rho_{*}\leq\rho_{l}$.
\item Finding new solution, we relaxed $\phi(r,t) $ to be static and began finding solutions in the two cases, namely $(c,d)$,$(c',d')$.




\item When $\Lambda\neq 0$, we proved the following theorem:
{\emph No-go Theorem}:
{\it Mimetic gravity doesn't have Linet-Tian family of cosmic strings. The only possible solution is when $\Lambda=0$}.

\end{itemize}

The theorem in absence of cosmological constant $\Lambda$ in cylindrical solution, with which we concluded, formed the true \emph{no- go} theorem to the existence of Linet-Tian cylindrical cosmic strings in mimetic gravity, and would alone suffice to establish the claim of equivalence of the mimetic gravity as a minimal  disformal deformation of Einstein gravity  to the Einstein-gravity with massless scalar field among  mathematical complexities.

%
%

\begin{acknowledgements}
We thank the referee for good observations and kind guide- lines.
\end{acknowledgements}



\end{document}